# Effect of Unfolding on the Spectral Statistics of Adjacency Matrices of Complex Networks

Sherif M. Abuelenin[a,b] , Adel Y. Abul-Magd[b,c]

[a]Faculty of Engineering, Port Said University, Port Said, Egypt
[b]Faculty of Engineering Sciences, Sinai University, El Arish, Egypt
[c]Faculty of Science, Zagazig University, Zagazig, Egypt

**Abstract**

Random matrix theory is finding an increasing number of applications in the context of information theory and communication systems, especially in studying the properties of complex networks. Such properties include short-term and long-term correlation. We study the spectral fluctuations of the adjacency of networks using random-matrix theory. We consider the influence of the spectral unfolding, which is a necessary procedure to remove the secular properties of the spectrum, on different spectral statistics. We find that, while the spacing distribution of the eigenvalues shows little sensitivity to the unfolding method used, the spectral rigidity has greater sensitivity to unfolding.

*Keywords:* Random matrix theory; unfolding; adjacency matrix; network analysis.

## 1. Introduction

Random matrix theory (RMT) provides a suitable framework to describe quantum systems whose classical counterpart has a chaotic dynamics [1, 2]. It models a chaotic system by an ensemble of random Hamiltonian matrices that depends only on the symmetry properties of the system. Systems with time reversal invariance, for example, are modelled by the Gaussian orthogonal ensemble (GOE). Starting from the discovery of the Wigner semicircle law in nuclear spectra [3] RMT has have a wide range of applications from quantum chaos to irreversible classical dynamics and low density liquids [2]. Random matrices are finding an increasing number of applications in the context of electrical engineering and related fields, especially in information and communication technology. The earliest applications of RMT to wireless communication were the works of Foschini [4] and Telatar [5] on characterizing the capacity of multi-antenna fading channels. Other RMT applications in wireless networks including capacity evaluation in multi-dimensional networks are extensively discussed in [9]. Reference [6] introduces the using of random matrices in modelling various wireless communication channels and problems, and reviews some of the typical applications of RMT in wireless communications. Other applications include characterizing the capacity of communication networks [10], modelling multi–path propagation between a transmitting and receiving antenna array [11], and neural networks learning [12]. The authors of [8] suggest that RMT is useful in the study of electric grid networks, and provides a starting point to applications in real-world smart grids.

Whole information about a network is encoded in its adjacency matrix. For definiteness we consider in this paper undirected networks only. The adjacency matrix is symmetric with 0's on the diagonal and off-diagonal elements equal to 1 or 0 depending on whether the related nodes are connected or not. The number of 1's in a row follows a Gaussian distribution with mean $p$ and variance $p(1 − p)$. This type of matrix is very well studied within the RMT framework [13-15]. RMT aims to understand correlations between eigenvalues independently of the variation of level spacing. For this purpose, it is common to "unfold" the spectrum by means of a transformation [16] involving the cumulated level density, so that the mean level spacing is equal to one. In most of practical applications, the exact form of the cumulated level density is not known. Unfolding is done by arbitrarily parametrizing the numerically obtained level density in terms of a polynomial. This introduces ambiguities in the unfolding procedure. In the rest of the paper we explain the random matrix approach to studying networks, and then we study the effect of the unfolding procedure on the calculated spectral statistics of the considered networks.

## 2. Methodology

*2.1. Random matrix approach to network*

The use of RMT in analyzing complex networks is based on the concept of representing a network as an adjacency matrix. The adjacency matrix is an $N{\times}N$ matrix, where $N$ represents the number of nodes in a network, and its off-diagonal elements represent links in the network. The simplest case of an adjacency matrix is that of an un-weighted undirected network. In this case the adjacency matrix is symmetric matrix whose elements are either 0 or 1, where 1 represents a link between two nodes, and 0 represents no link present between the two nodes. One case of special interest occurs when studying the interconnections between two or more networks. The overall network formed in such case consists of clusters of densely connected networks with sparse connections between the clusters. The adjacency matrix in this case is a block-diagonal matrix. Increasing the number of interconnections changes the matrix gradually from the block diagonal form to the form of an adjacency matrix of a single randomly connected network. This last case is very interesting when studying networks because it aids in understanding the transition from two or more separate networks to one big network through the introduction of intra-networks connections. This case was used to identify new characteristics of electric grid networks [8], where RMT tools were used to describe the interconnection of multiple grids and construct a simple model of a distributed grid, showing the transition from Poisson statistics, an indicator of regularity, to that of a GOE.

*2.2. Unfolding*

As RMT is capable of making predictions for the fluctuations on the scale of the mean level spacing, one has to remove the influence of the level density by unfolding the spectra [16]. This is often done by calculating the cumulative spectral function as the number of levels below or at the level $E$. It is frequently referred to as staircase function. It can be separated into an average part $N_{ave}(E)$, whose derivative is the level density, and a fluctuating part $N_{fluc}(E)$. $N_{ave}(E)$ is calculated for the matrix by running spectral average. If one knows the functional form of the mean level density $\rho(E)$, one obtains

$$I(E) = \int_{-\infty}^{E} dE' \rho(E').$$

For example, if the adjacency matrix belongs to a GOE, the mean level density is given by Wigner's semi-circle law

$$\rho_{GOE}(N,E) = \begin{cases} \dfrac{2N}{\pi a^2}\sqrt{a^2 - E^2}, & \text{for } |E| \leq a \\ 0, & \text{for } |E| > a \end{cases}. \tag{1}$$

where $a$ is the radius of the semi-circle, which is related to the standard deviation $\sigma$ of the off-diagonal elements of the adjacency matrix by

$$a = 2\sigma\sqrt{N}$$

In this case,

$$I_{GOE}(N,E) = N\left[\frac{1}{2} + \frac{E}{\pi a^2}\sqrt{a^2 - E^2} + \frac{1}{\pi}\arctan\left(\frac{E}{\sqrt{a^2 - E^2}}\right)\right], \text{ for } |E| \leq a. \tag{2}$$

On the other hand, if the exact form of the mean level density is not known, one has to smooth the averaged staircase function or predict it in some way. Unless one knows the exact form of the level density, unfolding is not a unique procedure. A priori, there is no criterion whether the numerical estimated $I_{av}(E)$ is close to the real one or not. Most frequently, one fits $N_{ave}(E)$ to a polynomial of degree $n$, often with $n = 3$. After extraction of the average part $I_{av}(E)$, it is unfolded from the spectra by the introduction of a dimensionless variable

$$\varepsilon_i = I_{av}(E_i) \tag{3}$$

In this variable, the spectra have mean level spacing unity everywhere. Thereafter, the observables calculated for the unfolded spectrum for each matrix are averaged over the ensemble. This is in the same spirit as it was done in spectra of nuclei [17]. These spectra were unfolded for each nucleus separately.

2.3. Spectral statistics

In order to compare the random matrix model with the experimental data, we consider the nearest neighbor spacing distribution (NNSD) $P(s)$, where $s_i \equiv \varepsilon_{i+1} - \varepsilon_i$ is the spacing between neighboring eigenvalues after unfolding. In the GOE case, the distribution is well-approximated by the Wigner surmise

$$P(s) = \frac{\pi}{2}s\, e^{-\frac{\pi}{4}s^2}. \tag{4}$$

NNSD strongly depends the short range "interaction" between the levels. The statistical analysis of long-range correlations of level spectra is usually carries out in terms of the level number variance $\Sigma^2$ and the spectral rigidity $\Delta_3$ [1,2]. By definition, $\Sigma^2(L)$ measures the level number variance in an interval of length $L$ of the unfolded spectrum. For a GOE,

$$\Sigma_{GOE}^2(L) = \frac{1}{2\pi^2}[\ln(2\pi L) - \cos(2\pi L) - Ci(2\pi L) + \gamma + 1 + \frac{1}{2}Si^2(\pi L)]$$
$$- \frac{1}{\pi}Si(\pi L) + 2L(1 - \frac{2}{\pi})Si^2(2\pi L) \qquad (5.a)$$

where $Si(r)$ and $Ci(r)$ are the sine- and cosine-integral functions, respectively, and $\gamma$ is Euler's constant [18]. The Dyson-Mehta $\Delta_3(L)$ statistic, also known as spectral rigidity, is another quantity which is often used to characterize the long-range correlations in quantum spectra. This is a measure of the average deviation of the spectrum on a given length $L$ from a regular "picket fence" spectrum of a harmonic oscillator. The relationship between $\Delta_3(L)$ and $\Sigma^2(L)$ is given in [1, 2] as:

$$\Delta_3(L) = (2/L^4)\int_0^L (L^3 - 2L^2 r + r^3)\Sigma^2(r)dr \qquad (5.b)$$

## 3. Results

In the following we present results for NNSD and $\Delta_3$ statistic for different ensembles of adjacency matrices that model random and clustered networks. For all cases to be considered, we construct ensembles of 20 1000-dimensional matrices, calculate the spectral characteristics for each matrix separately, and then take the ensemble average for each of the characteristics.

### 3.1. Random networks

In random graph model of Erdös and Rényi, any two nodes are randomly connected with probability $p$ [14]. As explained above, the adjacency matrix **A** of the random graph is a real symmetric $N \times N$ matrix: $A_{ij} = A_{ji} = 1$, if vertices $i$ and $j$ are connected, or 0, if these two vertices are not connected. The term "uncorrelated random graph" for a graph is used if (i) the probability for any pair of the graph's vertices being connected is the same, $p$; (ii) these probabilities are independent variables. In the $N \to \infty$ limit, the rescaled spectral density of the uncorrelated random graph converges to the semi-circle law of Eq. (1) [19] although many of the semi-circle law's conditions do not hold, e.g., the expectation value of the entries is a non-zero constant: $p \neq 0$. Recently, Palla and Vattay [20] have shown that the statistical properties of random networks drastically change with the mean degree of nodes $k = pN$ near the critical point of the percolation transition at $k_{cr} = 1$. They show that before the critical transition network follows Poisson statistics and as average degree is increased $k \gg 1$ they show GOE statistics.

An illustration of the convergence of the average spectral density to the semi-circular distribution can be seen on Fig. 1. The figure compares the semicircle law density function (1) with the average of 20 empirical density functions of the eigenvalues of 1000×1000 adjacency matrices. Three figures are plotted for $\rho(E)$ with $p = 0.001$, 0.01 and 0.1, which correspond to $k = 1$, 10 and 100, respectively. In order to keep figures simple, for the spectral density plots we have chosen to show the spectral density on single figure, and to rescale the horizontal ($E$) and vertical ($\rho$) axes by $\sigma^{-1}N^{-1/2} = [Np(1-p)]^{-1/2}$ and $\sigma N^{1/2} = [Np(1-p)]^{1/2}$, respectively.

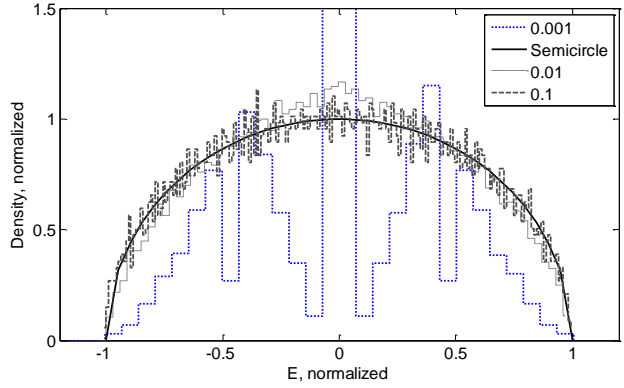

Fig. 1. Average spectral density compared to the semicircle law density function for p=0.001, 0.01, and 0.1. Horizontal and vertical axes (E, and ρ) are rescaled by by $\sigma^{-1}N^{-1/2} = [Np(1-p)]^{-1/2}$ and $\sigma N^{1/2} = [Np(1-p)]^{1/2}$, respectively.

In the following, we shall consider the spectral fluctuations for random networks with connectivity $p = 0.1$. As follows from Fig. 1, the "exact" spectral density of each of these networks is well represented by the semi-circle formula (1). Therefore, unfolding of the eigenvalues of the adjacent matrices can be done in terms of the corresponding cumulative spectral density (2) using Eq. (3). Figures 2 and 3 compare NNSD and $\Delta_3$ statistics for each of the two cases with the results of RMT as given by Eqs. (4, 5).

RMT tells that $\rho(E)$ is a semicircle for large GOE matrices. There are few other systems for which $\rho(E)$ is known. However, there is no natural choice for $\rho(E)$ in many systems. In this case unfolding is done by approximating the empirical staircase function by a polynomial of degree n:

$$I_{av}(E) = \sum_{i=1}^{n} c_i E^i$$

In Fig. 2 (a) NNSD $P(s)$ are shown by histograms for ensembles of 1000×1000 adjacency matrices for random networks with connectivity $p = 0.1$. In calculating the solid histograms, unfolding was done with $I_{av}(E)$ approximated by the GOE formula in Eq. (2). The rest of the histograms are for unfolding carried out by approximating Iav(E) with a 3rd, 4th, and 5th order polynomials. The line marked with diamonds is calculated the theoretical GOE formula of Eq. (4). Fig. 2 (b) shows the spectral rigidity $\Delta_3(L)$ for the same ensemble of adjacency matrices for random networks with connectivity: $p = 0.1$. In calculating the solid dots, unfolding was done with $I_{av}(E)$ approximated by the GOE formula in Eq. (2). The open dots, squares and diamonds are for unfolding carried out by approximating $I_{av}(E)$ with a third-, forth- or fifth-order polynomial. The solid line is calculated the GOE formula of Eq. (5).

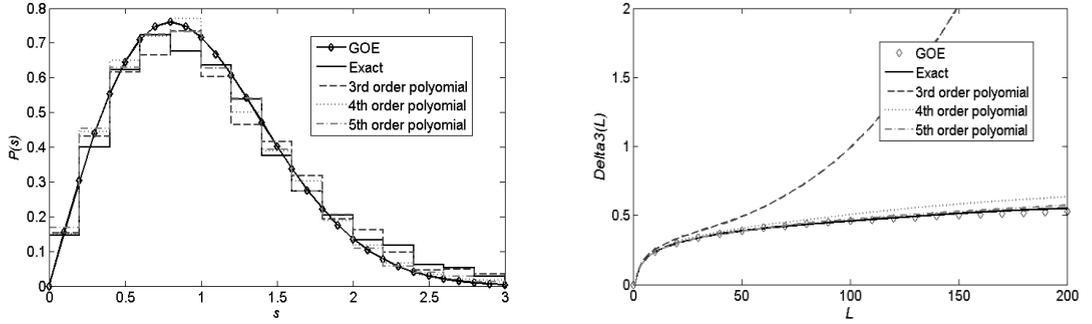

Fig. 2. (a) NNSD $P(s)$ for random networks with connectivity p = 0.1; (b) The spectral rigidity $\Delta_3(L)$ is shown for the same ensemble of adjacency matrices

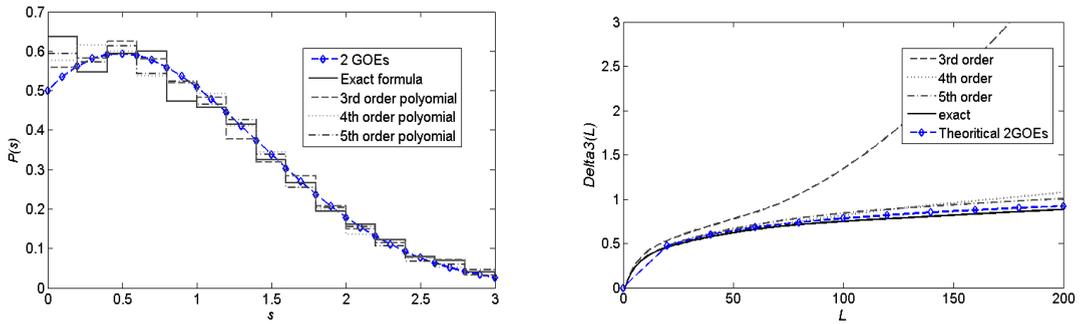

Fig. 3. (a) NNSD $P(s)$ for network composed of two unconnected clusters with connectivity p = 0.1; (b) The spectral rigidity $\Delta_3(L)$ is shown for the same ensemble of adjacency matrices

### 3.2. Clustered networks

Nodes that are heavily interconnected form a cluster or community. The nodes are reordered to display the blocks down the diagonal of the adjacency matrix representing the network. The adjacency matrix in this case is in block-diagonal form. NNSD is shown in Fig. 3 (a) by histograms for ensemble of 20 1000×1000 block-diagonal adjacency matrices for networks composed of two clusters with connectivity $p$ = 0.1. In calculating the solid histograms, unfolding was done with $I_{av}(E)$ approximated by the GOE formula in Eq. (2). The rest of the histograms are for unfolding carried out by approximating $I_{av}(E)$ with a $3^{rd}$, $4^{th}$, and $5^{th}$ order polynomials. The line marked with diamonds is calculated the theoretical super position of 2 GOE formulas. Fig. 3 (b) shows the spectral rigidity calculated for the same cases.

### 4. Conclusions

Whole information about a network is encoded in its adjacency matrix. Studying the eigenvalues of the adjacency matrix of a complex network provides useful information on the network characteristics. We studied the effect of the spectral unfolding, which is a necessary procedure to remove the secular properties of the spectrum of the adjacency matrix of networks, on different spectral statistics. We find that, while the spacing distribution of the eigenvalues shows little sensitivity to the unfolding method used, the spectral rigidity (and level number variance, since they are directly related), is very sensitive to unfolding. Since there are few systems for which $\rho(E)$ is known, exact unfolding cannot be performed in a systematic way for any system under consideration. This means that extra caution should be taken when examining the long-term correlation properties of real complex networks using RMT analysis.